\documentclass[11pt]{article}
\usepackage{moriond,epsfig}

\def\be{\begin{equation}}
\def\ee{\end{equation}}
\def\bea{\begin{eqnarray}}
\def\eea{\end{eqnarray}}

\begin{document}
\vspace*{4cm}
\title{Cosmological Brane Perturbations
\footnote{Based on work with Helen Bridgman and David Wands of the ICG, 
Portsmouth, and David Langlois and  
Mar\'{\i}a Rodr\'{\i}guez-Mart\'{\i}nez of GRECO, Paris.}
}
\author{ Karim A. Malik}
\address{GRECO, Institut d'Astrophysique de Paris, C.N.R.S.,
98bis Boulevard Arago, \\
75014 Paris, France}
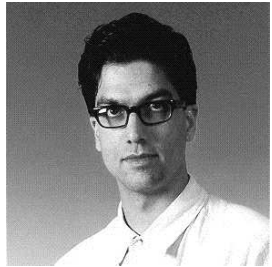
\maketitle
\abstracts{
Two approaches to the study of cosmological perturbations in the
brane-world scenario are compared:  
the first uses the 5D equations directly whereas the second approach   
projects them onto the 4D brane and then uses the effective 4D
equations. 
}
%
\section{Introduction}

At the end of the last millennium it was realized that the
traditional Kaluza-Klein approach is not the only possibility of
dimensional reduction. An alternative approach was developed, the
brane-world scenario. 
In this scenario the standard matter fields are constrained to 
a lower dimensional
hypersurface, or brane, which is embedded in a higher dimensional
spacetime, or bulk. The gravitational field is not restricted to the
brane and permeates the bulk as well as the brane \cite{BDL,RS2,BDEL}.

One of the goals of current brane-world research is to contrast
theoretical predictions with observations.  In standard 4D cosmology
\cite{LLBook} one distinctive feature of the early universe models is
the spectrum of density perturbation these models predict. 
In higher-dimensional models of the universe the evolution of
cosmological scalar perturbations and the possible observational
consequences these models have is still an open issue and a lot of
work is currently devoted to calculating the spectrum of density
perturbations in the brane-world scenario.

In this note we shall restrict the discussion to scalar perturbations,
that is to perturbations that transform like scalars on 3-spaces of 
constant time.
Tensor perturbations are discussed in reference \cite{LMW} 
and vector perturbations e.g.~in \cite{BMW1}.

In the next section we will briefly outline the simple brane-world 
model we are investigating.
For a more thorough introduction to the brane-world scenario see 
for example the contribution of D.~Langlois to these 
proceedings \cite{DLMor}. In Sections \ref{4Dsec} and \ref{5Dsec}
we shall describe two different approaches to solve the 
cosmological perturbation equations that arise from the brane-world 
model. We finish with a brief conclusion.

\section{The Brane-world}

We will assume that our 4-dimensional world is described by a 3-brane
embedded in a 5-dimensional bulk.
The equations of motion in the brane-world are given by Einstein's
equations
\be
\label{5DEin}
G_{AB}=\kappa^2_5\left(-\Lambda_5g_{AB}+T_{AB} \right) \,,
\ee
where $G_{AB}$ is the 5D Einstein tensor, $g_{AB}$ is the 5D metric,
$\Lambda_5$ is the bulk cosmological constant, and $T_{AB}$ is the
energy momentum tensor of the bulk fields. 
The standard matter fields that live on the brane, 
enter the picture only through
the Israel junction conditions: these junction conditions
relate the matter on the brane, described by
the energy momentum tensor $T_{\mu\nu}$ and the brane tension
$\lambda$, to the extrinsic curvature $K_{\mu\nu}$ of the brane, and
are given by \cite{israel,SMS}
\be
\label{junc}
K_{\mu\nu}
=-\frac{\kappa_5^2}{2}\left(
T_{\mu\nu}-\frac{1}{3}g_{\mu\nu}(T-\lambda)\right) \,,
\ee
where $g_{\mu\nu}$ is the 4D metric on the brane and  we assumed that the
brane is located at a fixed point in the $Z_2$ symmetric bulk.

To solve the Einstein equations (\ref{5DEin}) subject to the junction
conditions Eq.~(\ref{junc}) we have two options: either to use the 5D
equations directly or to project everything onto the brane and then use
the ensuing 4D equations. We shall describe the latter in the next
section and the former in Section \ref{5Dsec}.

\section{The Projective 4D Approach}
\label{4Dsec}

For a 4D observer it is natural to ask, what he or she is able to observe
living in a 5D brane-world but being restricted to the 4D brane.
Using the Gauss and the Codazzi equations Shiromizu et.al.~\cite{SMS}
showed that the 5D Einstein equations (\ref{5DEin}) can be projected
onto the brane to give the effective four-dimensional Einstein
equations
\footnote{ Note, that in this section we have set the bulk energy
momentum tensor $T_{AB}=0$, since life is difficult enough without it.
Here, the only energy in the bulk is the 5D cosmological constant.}
\be
\label{4DEin}
{}^{(4)}G_{\mu\nu}+\Lambda_4 g_{\mu\nu}
 = \kappa_4^2 T_{\mu\nu}+\kappa_5^4\, \Pi_{\mu\nu} -E_{\mu\nu}\,,
\ee
where the effective cosmological constant on the brane is 
$\Lambda_4 =\frac{1}{2} \Lambda_5 +\frac{\kappa_5^4}{12}\,\lambda^2$,
the 4D coupling constant is related to the 5D coupling constant by
$\kappa_4^2 = {\kappa_5^4\,\over6}\lambda$ 
and $E_{\mu\nu}$ is the projected 5D Weyl tensor, which describes
the non-local effect of the gravitational field. It is defined as
\begin{equation}
E_{\mu\nu} \equiv C^E_{~A F B}n_E n^F g_\mu^{~A}
g_\nu^{~B} \,,
\end{equation}
where $n_A$ is the normal vector to the brane. 
The projected Weyl tensor $E_{\mu\nu}$ acts like an imperfect radiation
fluid with anisotropic stress \cite{LMSW,BMW2}. 
The quadratic energy momentum tensor $\Pi_{\mu\nu}$ 
is given by
\be
\Pi_{\mu\nu} = -\frac{1}{4} T_{\mu\alpha} T_\nu^{~\alpha}
+\frac{1}{12} T T_{\mu\nu}
+\frac{1}{8}g_{\mu\nu} T_{\alpha\beta} T^{\alpha\beta}-\frac{1}{24}
g_{\mu\nu} T^2\,. 
\ee
%
%
Although we started with the 5D Einstein equations we get 4D effective
equations which are independent of the evolution of the bulk
spacetime, being given entirely in terms of quantities defined on, or
near, the brane.

Unfortunately this leaves terms which are not completely determined by
the local dynamics on the brane~\cite{SMS,Roy}: As in the 4D case we
would like to define a quantity that can source the large scale CMB 
anisotropies.
Conservation of energy on the brane allows us to construct a curvature
perturbation on uniform density hypersurfaces, $\zeta$, which is
conserved on large scales for adiabatic matter perturbations
\cite{WMLL}.  In a similar vain we can construct a curvature
perturbation on uniform Weyl-fluid energy density hypersurfaces,
$\zeta_{\rm{Weyl}}$, since the Weyl-fluid energy density is also
conserved to linear order.  However, the quantity that sources the CMB
anisotropies is the total curvature perturbation $\zeta_{\rm{total}}$
and its evolution depends not only on $\zeta$ and 
$\zeta_{\rm{Weyl}}$ but also on terms
which can not be calculated using
quantities defined solely on the brane \cite{LMSW,BMW2}.

Nonetheless the dynamics and effective gravity on the
brane can be interpreted, and often most easily understood, in terms
of the effective four-dimensional Einstein equations.

\section{Using the Full 5D Equations}
\label{5Dsec}

So instead of using the projective approach described in the last
section we could use the full five-dimensional equations of motion
(\ref{5DEin}). This approach has enjoyed considerable
attention \cite{Roy,Langlois,cvdb,Mukoh,LMSW,Deruelle,Koyama,ivo,Kodama,Bozza,BMW2,BMW1}. 
For concreteness let us start with the metric
\begin{equation}
\label{pertmetric} 
g_{AB}= \left(
\begin{array}{ccc}
-n^2(1+2A) & a^2 B_{,i} & nA_y \nonumber\\ a^2 B_{,j} 
 & a^2\left[ (1+2{\cal{R}})\delta_{ij} + 2E_{,ij} \right] & a^2 B_{y,i}
 \nonumber\\
nA_y & a^2 B_{y,i} & b^2(1+2A_{yy}) \nonumber
\end{array}
\right) \, .
\end{equation}
Here $n$, $a$, and $b$ are scale factors and functions of coordinate
time $t$ and the extra dimension $y$, and $A,B,E,{\cal{R}},B_{y}, A_y$,
and $A_{yy}$ are the scalar metric perturbations and functions of $x^A$.
Although it is tedious, with the help of a computer algebra
package it is possible to write down the full 5D equations of motion.

One problem in solving the equations of motion is the sheer size of
the equations: Whereas the perturbed part of for example the 0-0
component of the 4D Einstein tensor has four terms in an arbitrary
gauge, the same component has 14 terms for the 5D metric given in
Eq.~(\ref{pertmetric}) \cite{LMR}.

Another, even more unpleasant problem is as follows:  
In standard 4D cosmology we Fourier decompose the perturbations
on 3-spaces of constant time, which reduces the equations of motion
to a system of ordinary differential equations in $t$ \cite{Bardeen}. 
In 5D, in order to be able
to relate the results of the calculations to observations, 
we decompose the metric also on spatial 3-spaces into Fourier modes.
But here we get
a system of coupled partial differential equations in $t$ and $y$. 
As one can imagine, this makes
finding general solutions extremely difficult.

Nevertheless progress has been made.
A lot of work has been devoted to develop the formalism, which
is by now quite well understood.
In order to be able to rewrite the problem in terms of ordinary
differential equations quite often a simplified background is used.
For example if we choose the background to be static Minkowski space, 
we assume that the scale factors reduce to $n=b=a$ where
$a=a(y)$ and also any other background quantities are $y$-dependent only. 
Although this is quite a severe simplification, it
allows for considerable progress since the equations of motion
are now more likely to decouple,
see e.g.~\cite{Bozza} in the particular case of a dilatonic 
brane-world model.

\section{Conclusion}
\label{conc}

The projective approach, described above in Section \ref{4Dsec} gives 
useful physical insights into the physics of the brane-world scenario.
Unfortunately it does not give, in general, a closed system of
equations.
We therefore advocate the use of the full 5D equations, as described 
in the last section, as the way forward. More analytical work needs
to be done to solve the evolution equations in a general setting,
probably accompanied by numerical efforts.

Progress so far has been slow and painful for cosmological brane
perturbations, but it is still early days \ldots

\section*{Acknowledgments}

I would like to thank the organisers for
a enjoyable and stimulating conference. 

This research has been supported by a Marie Curie Fellowship
of the European Community Programme
``\emph{Improving Human Research Potential and the Socio-economic
Knowledge Base}'' 
under the contract number \emph{HPMF-CT-2000-00981}.

\section*{References}

\end{document}